# How Software Development Group Leaders Influence Team Members' Innovative Behavior

Fabio Q. B. da Silva, Cleviton V. F. Monteiro, Igor Ebrahim dos Santos, Luiz Fernando Capretz

**INNOVATIVE BEHAVIOUR** is the behaviour exhibited by individuals that engage in the intentional generation, promotion, and realization of new ideas within a work role, workgroup, or organization in order to benefit role performance, the group, or the organization[1]. Innovative behaviour is not the same as innovation, although both concepts are closely related. Innovation is the implementation of a new or significantly improved product, service, process, business model, or organizational structure. For innovation to happen several ideas must be generated and the best ones selected to be developed and then deployed or marketed. The generation of new ideas and their promotion and realization in the workplace are the results of individuals expressing their innovative behaviour.

In our studies of industrial practice, we gathered several examples of such behaviour. For instance, in a software company case study, we observed a software engineer who created and implemented a new process in a manufacturing support system that cut the time consumed in tool calibration from 30 to 3 minutes, considerably reducing the downtime of the production line for all clients of the system. What is relevant in this example is that the software engineer took the initiative to develop the new process during her spare time (it was not part of her duties), then promoted the new idea with her project manager, and finally presented the idea to the clients to make its implementation viable.

Being a human behaviour expressed in a social environment, innovative behaviour is likely to be affected by a diverse and complex network of factors at organizational, workgroup, and individual levels. In particular, in a workgroup context, group leaders exercise (consciously or not) their influence in ways that may increase or decrease the "likelihood of idea generation by followers and the subsequent development of these ideas into useful products"[2].

Therefore, the importance of the innovative behaviour of software engineers to promote innovation in industry motivated us to investigate how software team leaders (project managers, Scrum masters, technical leaders, etc.) might influence the innovative behaviour of team members. The empirical evidence supporting the claims we make below come from the synthesis of findings from two sources. First, a systematic literature review conducted in 2013 that analysed 80 articles covering the period of 1964-2012 (see Box 1). Second, two industrial case studies performed in software companies in Brazil and Canada, which were conducted between November 2012 and March 2014 with the participation of 76 software engineers.

Systematic Literature Review Details (Box 1)

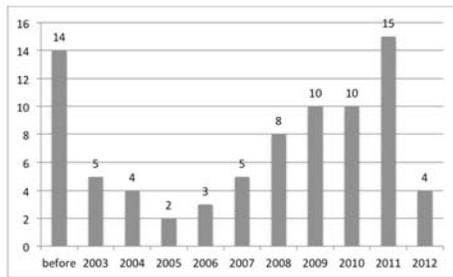
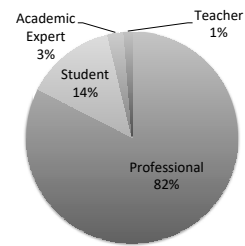

Types of Study Subjects

A total of 60 constructs were mapped. The most studied leadership styles were: transformational leadership, transactional leadership, and charismatic leadership. The studies investigate the influence of these factors on the creativity and several individual behaviours, including innovative behaviour.

Innovative Behaviour and Creativity (Box 2)

Creativity, the generation of new and useful ideas[3], is an important component of innovative behaviour. It is closely related to the first stages of the innovation process. However, the striking difference between creativity and innovative behaviour is that for the latter the individual has also to promote and implement the idea until it is perceived as being useful as a new business component: product, service, process, etc.

**Leadership style and innovative behaviour**

Leadership Styles (Box 3)

*A transactional leader*[4] builds the foundation for relationships with followers in terms of clarifying responsibilities, specifying expectations and tasks requirements, negotiating contracts, and providing recognition and rewards in exchange for the expected performance.

A *transformational leader* "raises associates' level of awareness of the importance of achieving valued outcomes and the strategies for reaching them"[4]. They also encourage followers to transcend their self-interest for the sake of the team or organization, and to seek higher levels of achievement, autonomy, and affiliation, which can be both work related and not work related.

*An ambidextrous* leader has "the ability to foster both explorative and exploitative behaviors in followers by increasing or reducing variance in their behavior and flexibly switching between those behaviors."[5]

Several studies have investigated the relationship between leadership styles and creativity (See Box 2), innovation, and performance. According to these studies, transactional and transformational leadership are distinct styles of leadership that are likely to have different effects on follower's (innovative) behaviour (See Box 3).

Transformational leadership has been associated with attributes such as inspirational motivation and intellectual stimulation, whereas transactional leaders are associated with practices such as contingent reward and management-by-exception. The characteristics and attributes of these two leadership styles made researchers to hypothesize that transformational leadership would create better conditions for the followers to express innovative behaviour.

In our literature review, studies have reported a positive relationship between transformational leadership and creativity. Further, there are also evidences to support that this relationship is moderated by individual psychological empowerment, group knowledge sharing, and collective efficacy. Other studies in the review identified a negative relationship between transactional leadership and creativity. In particular, this effect happens through its negative influence on collective efficacy and knowledge sharing.

Yet, other researchers also showed positive relationship of transactional leadership and exploitative innovation, whereas transformational leadership was related to exploratory, or radical, innovation[6]. Environmental dynamism, or the rate of change in the preferences of consumers and the products of the organization, is a key moderator to understand the balance between the two types of innovation, and hence the balance between the styles of leadership in promoting innovative behaviour.

Finally, some investigators also found negative relationship between transformational and transactional leadership and creativity or innovation, showing that context plays a significant role in explaining research findings[7].

It has become evident that variation among research findings indicates that the traditionally studied leadership styles are too broad in nature to promote innovation as they might both foster and hinder innovation under specific circumstances. Therefore, it seems that to explain the relationship between leadership and innovation, theories should incorporate behaviour flexibility instead of defining stable and inflexible behaviours. Indeed, innovation requires explorative and exploitative activities. Therefore, leaders should be able to behave in transactional and transformational ways so they can influence followers to engage in exploration and exploitation, as needed. This flexible style of leadership has been called in the literature ambidextrous leadership.

**What we know**

Integrating evidence from the literature with the findings of two case studies performed in Brazil and Canada, we arrive at four main results.

*Leader's acceptance of new ideas is positively related to followers' perception about the group acceptance of innovative behaviour.*

Group acceptance positively influences individual behaviour, in general. In the case of innovative behaviour, group acceptance of new ideas and of changes in the way the work is usually performed is positively related to innovative behaviour of group members. In a workgroup, leaders have an important role in the development of group values and behaviour acceptance. Therefore, the perception of group members about the group acceptance of innovative behaviour is positively related to the leader openness to new ideas coming from the group. That is, when the leader accepts new ideas and supports their development, the individual perception of group acceptance will be positive. However, when the leader avoids changes and does not support the development of new ideas, the perception will be that the workgroup is not open to ideas.

The acceptance of and support to new ideas must be exercised with care, as a balance must be found between getting the job done by sticking to the plans (usually less innovation) or constantly adding innovations at the potential risk of not delivering timely results. In our observations, ambidextrous leaders are effective in working with this balance.

*Leader proximity is positively related to the perception of followers about the group acceptance of innovative behaviour.*

The individual perceives more space to propose ideas when there is a closer relationship between the group and its leader. In turn, when the leader is not close, the individual is inhibited to propose ideas because both leader and follower spend a short time together and this time usually is used to perform tasks previously planned, which leaves little room for innovations. In addition, the individual filters ideas due to the fear of proposing something wrong or useless. Therefore, the channel to discuss new ideas is more effective as the leader works closer to the followers.

In our studies, we observed that transactional leaders tend to manage the tasks very close, being in constant contact with the group, whereas transformational leaders delegate more and, thus, tend to manage from the distance.

*Leadership support is positively related to followers' innovative behaviour.*

Workgroup leaders are usually responsible for the major decisions related to project planning. In addition, they are usually the technical or managerial reference for the group members. Therefore, the support they provide to the individuals is important to help overcoming the challenges as well as to get resources (e.g. time, equipment, software, and literature) to search for a good solution or to implement the ideas proposed.

Transactional leaders pay close attention to deviation from plans, including the use of resources. This leadership style is less prone to be flexible with unplanned requests for resources, which are usually necessary in developing new ideas. In addition, transactional leaders deal with deviations using hard criticism, which can result in followers taking the leader's desired, and therefore supported, pathway of approaching problems instead of trying new ways.

A*mbidextrous leadership strategies positively influence the followers' innovative behaviour.*

Our findings related to leadership style in software development agree with evidences from the literature of other fields that both leadership styles are needed to support innovative behaviour. Transactional leaders tend to manage the tasks more closely, giving the followers the perception of greater overall support to their activities. However, these leaders tend to be less flexible with resource management, reducing the actual support to unplanned activities. On the other hand, transformational leaders are likely to build the perception of managing from the distance, but at the same time can be more supportive for new and unplanned resource acquisition. The combination of styles, characteristic of ambidextrous leaders, offers the needed balance that can stimulate software developers to exhibit innovative behaviour in practice.

### *What are the practical implications?*

What we currently know about the influence of leaders on followers innovative behaviour shows that balance between styles of leadership will create important workplace conditions to foster the generation and promotion of ideas. In particular, leaders are likely to provide better support to their followers if they:

- Develop and use practices to listen to new ideas, thus creating the perception on followers that innovative behaviour is positively valued in the organization and in the group.
- Timely assess the viability of engaging resources for the development or refinement of new ideas, thus providing appropriate feedback on whether or not the idea will be further developed.
- Balance delegation, autonomy, and flexibility with close management of tasks and resources (because leader proximity increases the perception of space to propose ideas), being positively related to the followers' perception about group acceptance of innovative behaviour.

Evidence in the literature from several business sectors shows that exploratory and exploitative innovation strategies are complementarily important for competitiveness. Our empirical findings reinforced those evidences in the context of software development companies. The innovative behaviour of individuals is an essential ingredient to success in both types of innovations strategies and leaders can have a big influence on this behaviour. Adopting a leadership style that combines transactional and transformational practices is more likely to produce effective results in supporting innovative behaviour. In software development, project managers and other group leaders should be stimulated and supported in adopting such practices to create the conditions for innovative behaviour to thrive.

*FABIO Q. B. DA SILVA is an associate professor at Universidade Federal de Pernambuco in Brazil, where he leads the Human Aspects in Software Engineering research group (www.haseresearch.com). He also has extensive research work on empirical software engineering methods and evidence based software engineering. Contact him at fabio@cin.ufpe.br.*

*CLEVITON V. F. MONTEIRO is an adjunct professor at Universidade Federal Rural de Pernambuco in Brazil. As a member of the Human Aspects in Software Engineering research group his work is focused on innovation, innovative behaviour and motivation of software professionals. Contact him at cleviton@gmail.com.*

*IGOR EBRAHIM DOS SANTOS is a project manager in a multinational consulting company, where he has been leading groups of software engineers. He is also a member of the Human Aspects in Software Engineering research group. Contact him at ies@cin.ufpe.br.*

*LUIZ FERNANDO CAPRETZ is a professor of software engineering and assistant dean (IT & e-Learning) at Western University in Canada, where he also directed a fully accredited software engineering program. He has vast experience in the engineering of software and is a licensed professional engineer in Ontario. Contact him at lcapretz@uwo.ca or via [http://www.eng.uwo.ca/electrical/faculty/capretz_l/](http://www.eng.uwo.ca/electrical/faculty/capretz_l/).*